\documentclass[prb,twocolumn,superscriptaddress,showpacs,amsmath,amssymb]{revtex4}
\usepackage{amsfonts}
\usepackage{bbm}
\usepackage{graphicx}
\usepackage{dcolumn}
\usepackage{bm}
\usepackage{subfigure}
\usepackage{color}

\begin{document}
\title{Helimagnetic Josephson diode effect}
\author{Qiang Cheng}
\email[]{chengqiang07@mails.ucas.ac.cn}
\affiliation{School of Science, Qingdao University of Technology, Qingdao, Shandong 266520, China}
\affiliation{International Center for Quantum Materials, School of Physics, Peking University, Beijing 100871, China}

\author{Yu-Chen Zhuang}
\affiliation{International Center for Quantum Materials, School of Physics, Peking University, Beijing 100871, China}

\author{Qing-Feng Sun}
\email[]{sunqf@pku.edu.cn}
\affiliation{International Center for Quantum Materials, School of Physics, Peking University, Beijing 100871, China}
\affiliation{Hefei National Laboratory, Hefei 230088, China}

\begin{abstract}
We study the Josephson diode effect in the one-dimensional superconductor/helimagnet/superconductor junctions using the Green's function method. For the spin-singlet $s$-wave pairing in superconductors, it is found that the necessary conditions for the Josephson diode effect are the nonzero chemical potential and the conical magnetic configuration in the helimagnet. The diode efficiency is strongly dependent on the chemical potential, chirality, tilt angle and exchange coupling in the helimagnet. The high efficiency close to $40\%$ can be obtained for specific parameter values. The sign of the diode efficiency can be tuned by changing the chirality, tilt angle, exchange coupling and chemical potential. The dependence of the diode efficiency on the number of supercells in the helimagnet is also investigated. The characteristics of the supercurrent nonreciprocity and diode efficiency in the junctions are clarified through the symmetry analysis and the energy band calculations. The diode effect for the spin-triplet $p$-wave pairing in superconductors is also discussed and the nonzero chemical potential is no longer a necessary condition for the Josephson diode effect due to the equal-spin Cooper pair-mediated transport in the $p$-wave junctions. These results provide a scheme for the Josephson diode effect without spin-orbit coupling, which possesses the potential applications in the design of dissipationless electronic devices.
\end{abstract}
\maketitle

\section{\label{sec1}Introduction}
Helimagnets (HMs) with chirality have attracted considerable interest in condensed matter physics due to their peculiar spin structure and the resultant novel physical phenomena. Anomalous electromagnetic, magnetoelectric and transport properties in HMs have been reported theoretically and experimentally\cite{Yokouchi1,Gao,Jiang,Zadorozhnyi,Kimoto,Ustinov,Zadorozhnyi2}. Especially, the nonreciprocal transport in HM draws special research interest. For example, the formation of the chiral magnetic order dramatically alters the electrical magnetochiral effect in HMs and an anomalous enhancement of resistance is observed in the chiral conical phase\cite{Aoki}. Nonreciprocal thermal transport has been demonstrated in the multiferroic HMs and the controllable thermal rectification using external fields has been predicted\cite{Hirokane}. The distinct contributions of spin fluctuation and band asymmetry to nonreciprocal resistivity in chiral magnets have been identified\cite{Nakamura}. Chirality-dependent nonreciprocal resistivity can be drastically enhanced near the helimagnetic transition field and several possible mechanisms for nonreciprocal transport are discussed\cite{Masuda}. In addition, the proximity-induced spin-triplet correlations and the generation of long-ranged spin-triplet pairs have been examined in structures with helimagnetic spin textures\cite{Bobkov,Spuri}.

Recently, nonreciprocal electric transport in $\alpha$-EuP$_3$ has been demonstrated experimentally. Based on first-principles band calculations, it has been found that band asymmetry driven by the conical spin structure under a magnetic field is responsible for the experimental observations\cite{Mayo}. Subsequently, a Yoshimori-type minimal model of an HM on a one-dimensional (1D) atomic chain was developed, which successfully explains the emergence of band asymmetry and nonreciprocal transport in centrosymmetric $\alpha$-EuP$_3$ without the Dzyaloshinskii-Moriya interaction\cite{Deaconu}. The model also captures the basic characteristics of unconventional $p$-wave magnets. Large magnetoresistance, spin filtering and anisotropic bulk spin conductivity have been evaluated using the two-dimensional version of the model\cite{Brekke}. The robustness of the realized $p$-wave magnetism under the Rashba spin-orbit interaction has been clarified based on the model on a 1D chain model\cite{Hodt}. In a similar 1D helimagnetic chain, nonreciprocal spin transport and its dependence on the chirality, period and cone angle have been systematically investigated\cite{Okumura}.

In this paper, we study the Josephson diode effect (JDE) in 1D helimagnetic Josephson junctions. JDE offers a pathway for designing nondissipative electronic devices and has been experimentally realized in van der Waals heterostructures\cite{HWu}. Recently, various theoretical and experimental schemes for JDE have been proposed in different types of junctions, including those in 1D structures\cite{Kopasov,Davydova,Tanaka,Kokkeler,Cheng1,YFSun,Maiani,Lu,Costa1,Hess,Hodt,Nikolic,WTLu,Mondal,
Schulz,Qi,CSun,Costa2,YFSun2,Patil,Mazur,Scharf,Ilic,addr1,Vakili,Cheng2,Debnath,YMao,Cayao,Nunchot,Legg,Ciaccia,
Misaki,Fominov,WTLiu}. For instance, JDE in curved nanowire junctions subjected to an external field has been reported when the wire undergoes a transition to a topologically nontrivial state\cite{Kopasov}. When a metallic nanowire is placed on two superconductor (SC) slabs, an applied in-plane magnetic field can induce a spatially modulated pairing potential in the wire, leading to JDE in short Josephson junctions\cite{Davydova}. Josephson junctions consisting of three 1D SCs can support JDE that is switchable via the magnetic configuration of the two barriers\cite{Hodt}. Electrostatic gate-tunable JDE has been experimentally observed in short InSb nanowire Josephson junctions under a magnetic field perpendicular to the wire\cite{Mazur}.  Topological Josephson junctions formed by the edges of quantum Hall insulators can support JDE that is strongly dependent on the magnetic field and temperature\cite{Scharf}. The influences of Andreev and Majorana bound states on JDE in nanowire-based junctions have also been clarified in the presence of a Zeeman field\cite{Cayao,Mondal}.

Here, we propose 1D SC/HM/SC Josephson junctions to realize JDE and study the properties of nonreciprocity in the current-phase difference relations (CPRs). For the spin-singlet $s$-wave pairing in SCs, the first necessary condition for JDE in our junctions is the formation of a conical magnetic configuration in the HM, corresponding to a tilt angle not equal to $m\pi$ and $(2m+1)\frac{\pi}{2}$, where $m$ is an integer. The second necessary condition for JDE is a finite chemical potential in the HM, which supports the finite-momentum pairing responsible for JDE. Once these conditions are satisfied, the Josephson diode efficiency shows a strong dependence on the chirality, tilt angle, exchange coupling, chemical potential and size (i.e., the number of supercells) of the HM. The sign of the diode efficiency is opposite for the different chiralities of the HM. The diode efficiency is antisymmetric with respect to the tilt angle. The sign of the diode efficiency can also be tuned by the exchange coupling and the chemical potential in the HM. The maximum diode efficiency can reach values close to $40\%$ for specific parameter choices. The symmetry relations satisfied by the CPRs and the diode efficiency are clarified through symmetry analysis and energy band structure calculations. For spin-triplet $p_x$-wave pairing in SCs, the condition for the diode effect no longer requires a finite chemical potential because equal-spin Cooper pairs-mediated transport becomes possible in the conical state of the HM. The realization of JDE for both the $s$-wave and $p_x$-wave pairing in our junctions does not require spin-orbit coupling, offering a promising route for designing nondissipative devices.

The paper is organized as follows. In Sec.\uppercase\expandafter{\romannumeral 2}, we present the model of our junctions and derive the expression for the Josephson current. In Sec.\uppercase\expandafter{\romannumeral 3}, we discuss the numerical results for the CPRs and diode efficiency, and analyze the symmetric or antisymmetric relations satisfied by them. Sec.\uppercase\expandafter{\romannumeral 4} concludes the paper. Details of the energy band calculations and the operator representation for symmetry analysis are provided in the Appendix.

\section{\label{sec2}Model and Formulation}
\begin{figure}[!htb]
\centerline{\includegraphics[width=1\columnwidth]{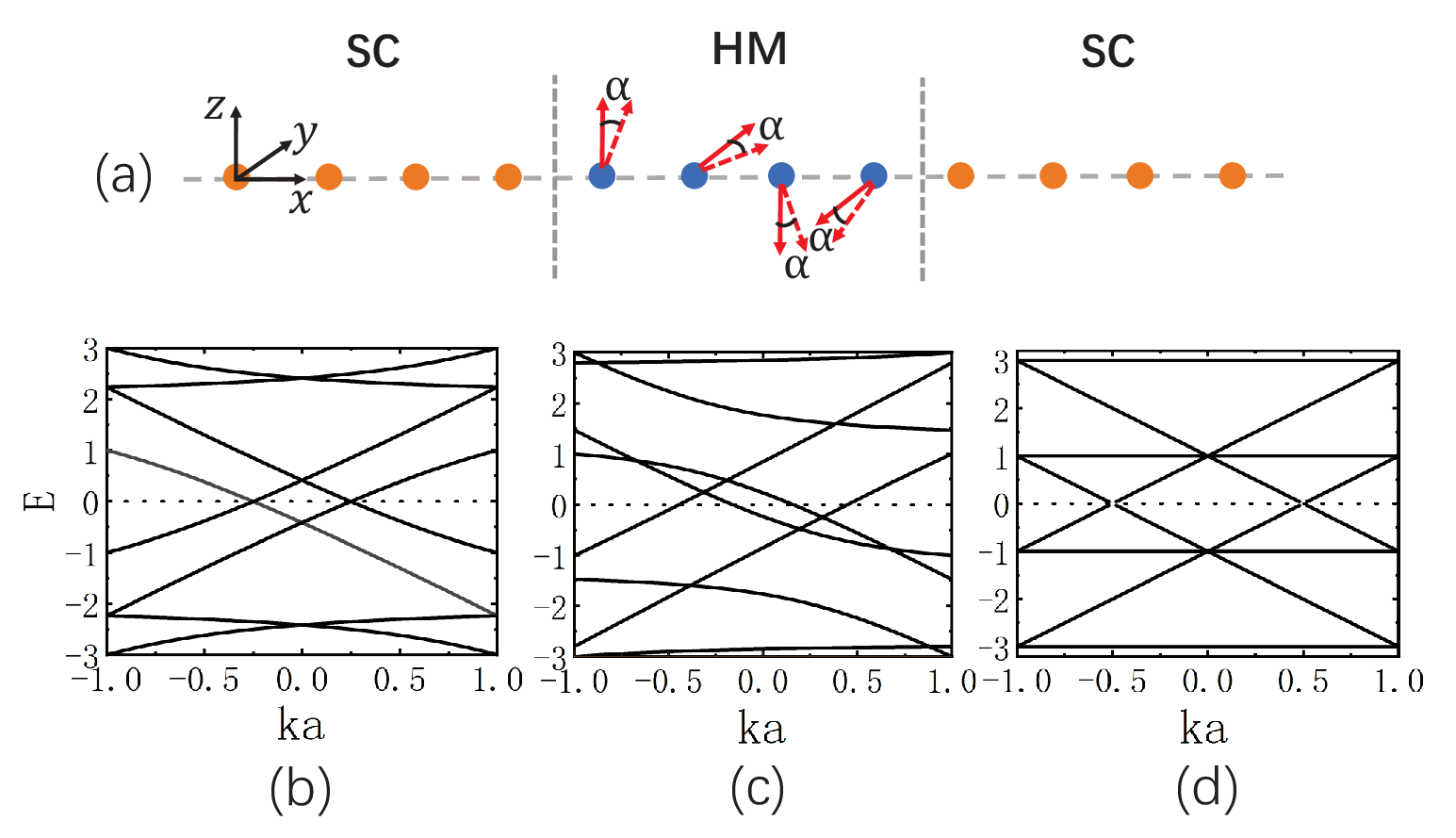}}
\caption{(a) Schematic illustration of the 1D SC/HM/SC Josephson junctions with one supercell in the HM. The gray dotted lines indicate the two interfaces of the junctions. The angle $\alpha$ describes the magnetic configuration of the HM under an external field. The solid red arrows represent the original helical magnetic configuration without tilt of the local moment. The dashed red arrows represent the conical magnetic configuration with a nonzero tilt angle $\alpha$.
(b-d) Energy bands of the HM for (b) $\alpha=0$ or $\pi$, (c) $\alpha=\frac{\pi}{4}$ and (d) $\alpha=\frac{\pi}{2}$ or $\frac{3\pi}{2}$, calculated with parameters $\lambda=1$,$M=1$,$t_0=1$, $\gamma=+1$ and $\mu=0$.}\label{fig1}
\end{figure}

The 1D SC/HM/SC Josephson junctions considered in this work are schematically shown in Fig.\ref{fig1}(a).
The original 1D HM chain consists of $N$ magnetic supercells, each containing four atoms. There is a $90^{\circ}$ rotation between the local magnetizations of the neighboring atoms, as indicated by the red solid arrows in Fig.\ref{fig1}(a). The Hamiltonian of the HM (represented by blue circles in Fig.\ref{fig1}(a)) under an external field can be written in spin space as\cite{Deaconu,Brekke,Hodt}
\begin{eqnarray}
\begin{split}
h=&-\sum_{i}t_0 c_{i}^{\dagger}c_{i+\delta x}+\text{H.c.}\\
&+\lambda\sum_{i}c_{i}^{\dagger}{\bf{M}}_{i}\cdot{\bf{\sigma}}c_{i}-\mu\sum_{i}c_{i}^{\dagger}c_{i},\label{hHM}
\end{split}
\end{eqnarray}
with $c_{i}=(c_{i\uparrow},c_{i\downarrow})^{T}$. Here, $t_0$ is the hopping amplitude between site $i$ and its nearest neighbor, $\lambda$ is the exchange coupling between electrons and the local magnetization ${\bf{M}}_{i}$ on site $i$, $\mu$ is the chemical potential and ${\bf{\sigma}}=(\sigma_x,\sigma_y,\sigma_z)$ are the Pauli matrices. The applied external field tilts the local magnetic moment by an angle $\alpha$, as shown in Fig.\ref{fig1}(a). The local magnetization under the field can then be expressed as
\begin{eqnarray}
{\bf{M}}_{i}=M\left[\sin{\alpha},\gamma\sin{\frac{i\pi}{2}}\cos{\alpha},\cos{\frac{i\pi}{2}}\cos{\alpha}\right].\label{M}
\end{eqnarray}
Here, $M$ is the magnitude of the local magnetization, $\gamma=\pm 1$ denotes the chirality and $i$ in Eq.(\ref{M}) indicates the position of atoms in the HM.

For simplicity, only one supercell containing four atoms in the HM is shown in Fig.\ref{fig1}(a). The positions of the four atomic sites are labeled $i=0,1,2,3$ from left to right. Next, we discuss the magnetic configuration of the HM as described by the tilt angle $\alpha$. First, we consider chirality $\gamma=+1$ in Eq.(\ref{M}). For $\alpha=0$, the local magnetizations on the four sites are in the $+z$, $+y$, $-z$ and $-y$ directions, respectively, as shown by the red solid arrows in Fig.\ref{fig1}(a). In this case, the HM exhibits a helical configuration of the local magnetization. For $\alpha=\pi$, the HM remains in a helical magnetic configuration, but the local magnetizations on the sites are inverted. The energy bands for $\alpha=0$ or $\pi$ are symmetric about $k=0$ as shown in Fig.\ref{fig1}(b). For $\alpha=\frac{\pi}{2}$, the local magnetization on each site is in the $+x$ direction and the HM becomes a ferromagnet with spin-polarization along the $+x$ direction. For $\alpha=\frac{3\pi}{2}$, the local magnetization on each site is in the $-x$ direction and HM turns into a ferromagnet with spin-polarization along the $-x$ direction. The energy bands for $\alpha=\frac{\pi}{2}$ or $\frac{3\pi}{2}$ are also symmetric about $k=0$ as shown in Fig.\ref{fig1}(d). For a general value of $\alpha$ not equal to $0,\frac{\pi}{2},\pi,\frac{3\pi}{2}$, the HM exhibits a conical magnetic configuration as depicted by the red dashed arrows in Fig.\ref{fig1}(a). The energy bands for such a general $\alpha$ are no longer symmetric about $k=0$ as shown in Fig.\ref{fig1}(c). Next, we consider chirality $\gamma=-1$ in Eq.(\ref{M}). For $\alpha=0$, the HM is in a helical magnetic state with the local magnetizations of the sites pointing in the $+z$, $-y$, $-z$ and $+y$ directions. For $\alpha=\pi$, the HM remains in a helical magnetic state with magnetizations opposite to those for $\alpha=0$. For $\alpha=\frac{\pi}{2}$ or $\alpha=\frac{3\pi}{2}$, the HM transforms into a ferromagnet with the same spin-polarization as for $\gamma=+1$.
For $\gamma=-1$, the shapes of energy bands in the HM for $\alpha=\frac{m\pi}{2}$
with $m$ being an integer are the same as those for $\gamma=+1$,
while the energy bands for the conical magnetic state
with $\alpha \ne \frac{m\pi}{2}$ satisfy
\begin{eqnarray}
E(\gamma=-1,k)=E(\gamma=+1,-k).\label{eb}
\end{eqnarray}
Details of the energy band calculations are provided in the Appendix \ref{secA}.

The Bogoliubov-de Gennes (BdG) Hamiltonian for the HM under an external field in the particle-hole$\otimes$spin space can be written as
\begin{eqnarray}
H=\sum_{i\in \text{HM}}\psi_{i}^{\dagger}H_{x}\psi_{i+\delta x}+\text{H.c.}+\sum_{i}\psi_{i}^{\dagger}H^i_{0}\psi_{i},\label{HMH}
\end{eqnarray}
with $\psi_{i}=(c_{i\uparrow},c_{i\downarrow},c_{i\uparrow}^{\dagger},c_{i\downarrow}^{\dagger})^{T}$. The matrices $H_{x}$ and $H^i_0$ are given by
\begin{eqnarray}
H_x=\text{diag}(-t_0,-t_0,t_0,t_0),\label{HMHx}
\end{eqnarray}
and
\begin{eqnarray}
H^i_0=\left(\begin{array}{cccc}
H^i_{011}&H^i_{012}&0&0\\
H^i_{021}&H^i_{022}&0&0\\
0&0&H^i_{033}&H^i_{034}\\
0&0&H^i_{043}&H^i_{044}\label{HMH0}
\end{array}\right),
\end{eqnarray}
with $H^i_{011}=\lambda M \cos\frac{i\pi}{2}\cos{\alpha}-\mu$, $H^i_{012}=\lambda M (\sin\alpha-j\gamma\sin\frac{i\pi}{2}\cos\alpha)$, $H^i_{021}=\lambda M (\sin\alpha+j\gamma\sin\frac{i\pi}{2}\cos\alpha)$, $H^i_{022}=-\lambda M \cos\frac{i\pi}{2}\cos{\alpha}-\mu$ and $H^i_{033}=-H^i_{011}$, $H^i_{044}=-H^i_{022}$, $H^i_{034}=-H_{012}^{i*}$, $H^i_{043}=-H_{021}^{i*}$. Note that we use $j$ (rather than $i$) to denote the imaginary unit.

For the left (right) semi-infinite SC (represented by orange circles in Fig.\ref{fig1}), the BdG Hamiltonian is
\begin{eqnarray}
\begin{split}
H_{L(R)}=&\sum_{i\in \text{SC}}\Psi_{L(R)i}^{\dagger}H_{L(R)x}\Psi_{L(R)i+\delta x}+\text{H.c.}\\
&+\sum_{i}\Psi_{L(R)i}^{\dagger}H_{L(R)0}\Psi_{L(R)i},\label{HLR}
\end{split}
\end{eqnarray}
with $\Psi_{L(R)i}=(c_{L(R)i\uparrow},c_{L(R)i\downarrow},c_{L(R)i\uparrow}^{\dagger},c_{L(R)i\downarrow}^{\dagger})^T$ in the left (right) SC. The matrices $H_{L(R)x}$ and $H_{L(R)0}$ are given by
\begin{eqnarray}
H_{L(R)x}=\text{diag}(-t_0,-t_0,t_0,t_0),\label{HLRx}
\end{eqnarray}
and
\begin{eqnarray}
\begin{split}
&H_{L(R)0}=\\
&\left(\begin{array}{cccc}
-\mu_{L(R)}&0&0&\Delta e^{j\phi_{L(R)}}\\
0&-\mu_{L(R)}&-\Delta e^{j\phi_{L(R)}}\\
0&-\Delta e^{-j\phi_{L(R)}}&\mu_{L(R)}&0\\
\Delta e^{-j\phi_{L(R)}}&0&0&\mu_{L(R)}
\end{array}\right),\label{HLR0}
\end{split}
\end{eqnarray}
where $\mu_{L(R)}$ is the chemical potential of the left (right) SC, $\Delta$ is the gap magnitude and $\phi_{L(R)}$ is the superconducting phase.
Here, we only consider the spin-singlet $s$-wave pairing in SCs. The case of the spin-triplet $p$-wave pairing will be discussed later.

The tunneling Hamiltonian can be given by
\begin{eqnarray}
H_T=\Psi_{Li=-1}^{\dagger}\check{T}\psi_{i=0}+\Psi_{Ri=4N}^{\dagger}\check{T}\psi_{i=4N-1}+\text{H.c}.
\end{eqnarray}
with $\check{T}=\text{diag}(t,t,-t,-t)$.
Here, the subscript $Li=-1$ denotes the rightmost site of the left SC, $i=0$ denotes the leftmost site of HM, $Ri=4N$ denotes the leftmost site of the right SC and $i=4N-1$ denotes the rightmost site of HM. The integer $N$ is the number of supercells in HM.

The particle number operator in the left SC can be written as
\begin{eqnarray}
N=\sum_{i\in \text{SC},\beta=\uparrow\downarrow}c_{Li\beta}^{\dagger}c_{Li\beta},
\end{eqnarray}
and the Josephson current in the junctions can be expressed as\cite{Cheng1,Cheng2,WTLiu,Cheng3,QFSun}
\begin{eqnarray}
\begin{split}
I=&e\langle\frac{dN}{dt} \rangle
= -\frac{e}{2\pi}\int dE\text{Tr}[\Gamma_z\check{T}G_{ML}^{<}+\text{H.c.}],
\end{split}
\end{eqnarray}
with $\Gamma_z=\sigma_z\otimes1_{2\otimes2}$. Here, the lesser Green's function $G_{ML}^{<}(E)$ can be written as $G_{ML}^{<}=-f(E)[G_{ML}^{r}(E)-G_{ML}^{a}(E)]$ with the Fermi distribution function $f(E)$. The retarded Green's function $G_{ML}^{r}$ and the advanced Green's function $G_{ML}^{a}$ can be derived from the matrices in Eqs.(\ref{HMHx}),(\ref{HMH0}),(\ref{HLRx}) and (\ref{HLR0})\cite{Cheng1,Cheng2,WTLiu,Cheng3,QFSun}.

\section{\label{sec3}Numerical results and discussions}

In our calculations, we will take the hopping amplitudes $t_0=t=1$, the chemical potentials $\mu_L=\mu_{R}=1$ and the magnetization magnitude $M=1$. The gap magnitude of SCs is taken as $\Delta=0.01$. The superconducting phase difference is defined as $\phi=\phi_L-\phi_R$.

\begin{figure}[!htb]
\centerline{\includegraphics[width=1\columnwidth]{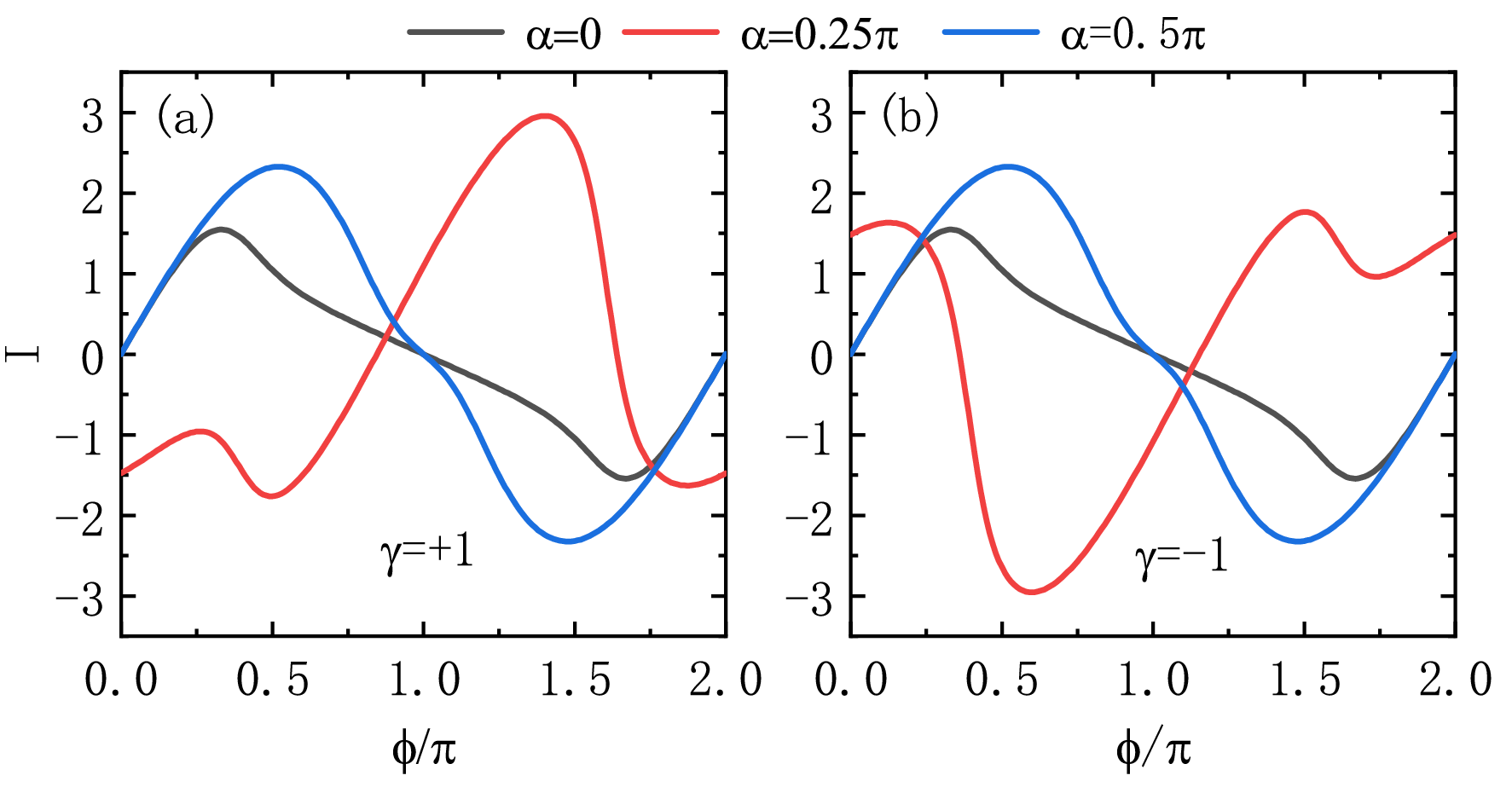}}
\caption{CPRs for different tilt angles with (a) $\gamma=+1$ and (b) $\gamma=-1$. Other parameters are taken as $\lambda=1$, $\mu=1$ and $N=1$.
}\label{fig2}
\end{figure}

In Fig.\ref{fig2}, we present CPRs for different values of the tilt angle $\alpha$. For chirality $\gamma=+1$ in Fig.\ref{fig2}(a), it is found that the Josephson current is reciprocal and JDE does not occur for $\alpha=0$. This can be demonstrated through symmetry analysis. We introduce the joint operation $X=M_{xz}TU_{1}(\frac{\pi}{2})$, where $M_{xz}$ is the mirror reflection about the $xz$ plane, $T$ is the time-reversal operation and $U_{1}(\frac{\pi}{2})$ is the gauge transformation.
Detailed results of applying the $M_{xz}$, $T$ and $U_{1}(\frac{\pi}{2})$
operations to the operator $c_{i\uparrow/i\downarrow}$ are provided in the Appendix \ref{secB}.
Under this joint operation, the Hamiltonian of the HM transforms as
\begin{eqnarray}
XH(\alpha)X^{-1}=H(-\alpha),\label{eqX}
\end{eqnarray}
and the Hamiltonian of the left (right) SC transforms as
\begin{eqnarray}
XH_{L(R)}(\phi_{L(R)})X^{-1}=H_{L(R)}(-\phi_{L(R)}),
\end{eqnarray}
which implies that the superconducting phase difference $\phi$ becomes $-\phi$. Since the time-reversal operation inverts the direction of the Josephson current, we obtain
\begin{eqnarray}
I(\alpha,\phi)=-I(-\alpha,-\phi).\label{sr1}
\end{eqnarray}
For $\alpha=0$, the relation in Eq.(\ref{sr1}) gives $I(\phi)=-I(-\phi)$ that ensures the critical currents along the $+x$ and $-x$ directions are equal. Therefore, JDE cannot occur for $\alpha=0$. In other words, JDE does not occur for the original helical magnetic configuration of the HM without the magnetization tilt, which is irrespective of the chirality $\gamma$. In this case, the CPRs are identified for $\gamma=\pm1$, as shown by the black lines in Fig.\ref{fig2}(a) and Fig.\ref{fig2}(b). The absence of JDE for the helical magnetic configuration of the HM is consistent with the symmetric energy bands about $k=0$ in Fig.\ref{fig1}(b).

For $\alpha=\frac{\pi}{2}$, JDE is also absent, as shown by the blue lines in Fig.\ref{fig2}(a) and Fig.\ref{fig2}(b). In this case, the HM becomes a ferromagnet with the local magnetization on each site along the $+x$ direction, independent of the chirality. The absence of JDE for $\alpha=\frac{\pi}{2}$ can also be demonstrated through symmetry analysis. We introduce the joint operation $Y=M_{yz}U_{1}(\frac{\pi}{2})$, where $M_{yz}$ is the mirror reflection about the $yz$ plane. The Hamiltonian of the HM is invariant under this operation while the Hamiltonian of the left (right) SC transforms as
\begin{eqnarray}
YH_{L(R)}(\phi_{L(R)})Y^{-1}=H_{L(R)}(\phi_{R(L)}).\label{Myztr}
\end{eqnarray}
The operation $M_{yz}$ exchanges the phases of the left and right SCs, thereby inverting the direction of the Josephson current. Consequently, we obtain
\begin{eqnarray}
I(\alpha=\frac{\pi}{2},\phi)=-I(\alpha=\frac{\pi}{2},-\phi),\label{srpi2}
\end{eqnarray}
which prohibits JDE from occurring for $\alpha=\frac{\pi}{2}$. The absence of JDE for the ferromagnetic configuration of the HM is consistent with the symmetric energy bands about $k=0$ in Fig.\ref{fig1}(d).

For $\alpha=0.25\pi$, JDE is clearly visible in the CPRs, as indicated by the red lines in Figs.\ref{fig2}(a) and (b). In this case, the HM exhibits the conical magnetic configuration. Therefore, we conclude that the necessary condition for JDE in our junctions is the formation of the conical magnetization under an external field. When this condition is satisfied, the relation $I(\phi)=-I(-\phi)$ no longer holds and JDE becomes possible.
From Fig.\ref{fig2}, we observe that for $\gamma=+1$ the critical current in the $+x$ direction exceeds that in the $-x$ direction, whereas for $\gamma=-1$, the opposite is true.
In fact, we have the relation
\begin{eqnarray}
I(\gamma,\phi)=-I(-\gamma,-\phi). \label{sr2}
\end{eqnarray}
We introduce the joint operation $Z=M_{xz}M_{yz}$. Under this operation, the Hamiltonian of the HM transforms as
\begin{eqnarray}
ZH(\gamma)Z^{-1}=H(-\gamma),\label{eqZ}
\end{eqnarray}
while the Hamiltonian of the left (right) SC transforms as
\begin{eqnarray}
ZH_{L(R)}(\phi_{L(R)})Z^{-1}=H_{L(R)}(\phi_{R(L)}).
\end{eqnarray}
In this joint operation, the reflection $M_{yz}$ exchanges the phases of the left and right SCs, thereby inverting the Josephson current. We then obtain the relation in Eq.(\ref{sr2}) which holds irrespective of the tilt angle $\alpha$. This relation expresses the effect of the chirality on the nonreciprocity of supercurrent in our junctions. Additionally, the presence of JDE for the conical magnetic configuration of the HM and its dependence on the chirality $\gamma$ can also be understood through the asymmetric energy bands in Fig.\ref{fig1}(c) and the relation in Eq.(\ref{eb}). Recently, the geometric diode effect in a kinked nanostrip with Rashba spin-orbit coupling is studied by Maiellaro et al\cite{Maiellaro}. Our symmetry analysis is consistent with the symmetric relations for the supercurrent obtained by these authors.

It is instructive to compare the CPRs results in our junctions with chirality to those obtained for the chiral quantum dot in Ref.[\onlinecite{Cheng1}]. From Fig.\ref{fig2}, we see that the realization of JDE is closely related to the appearance of a $\cos{\phi}$-type component in the CPRs. For $\alpha=0$ or $\frac{\pi}{2}$, $I(\phi=0)=I(\phi=\pi)=0$, indicating the absence of a $\cos{\phi}$-type current in the CPRs and consequently JDE does not occur. For $\alpha=0.25\pi$, $I(\phi=0)\ne0$ and $I(\phi=\pi)\ne0$ means the presence of a $\cos\phi$-type current in CPRs and JDE is realized. However, the realization of JDE in the SC/chiral quantum dot/SC junctions in Ref.[\onlinecite{Cheng1}] does not require the presence of a $\cos\phi$-type current in the CPRs.
\begin{figure}[!htb]
\centerline{\includegraphics[width=1\columnwidth]{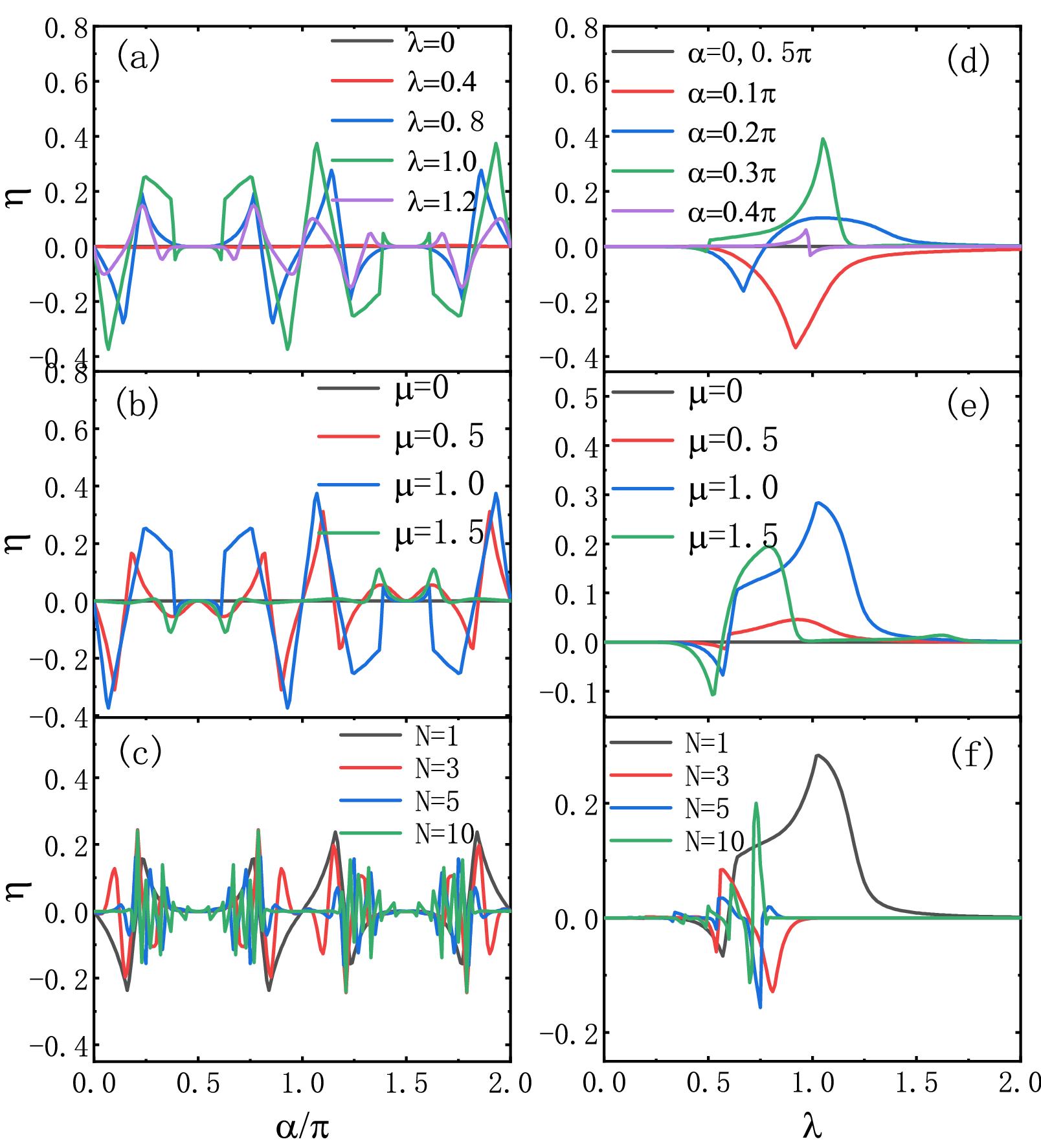}}
\caption{(a)-(c) The JDE efficiency as a function of the tilt angle. (d)-(f) The JDE efficiency as a function of the exchange coupling strength. The parameters are taken as (a) $N=1$ and $\mu=1$, (b) $N=1$ and $\lambda=1$, (c) $\lambda=0.75$ and $\mu=1$, (d) $N=1$ and $\mu=1$, (e) $N=1$ and $\alpha=0.25\pi$, and (f) $\alpha=0.25\pi$ and $\mu=1$.}\label{fig3}
\end{figure}

To quantitatively characterize JDE in our junctions, we define the JDE efficiency $\eta$ as
\begin{eqnarray}
\eta=\frac{I_{c+}-I_{c-}}{I_{c+}+I_{c-}}.
\end{eqnarray}
Here, the critical current in the $+x$ direction is defined as $I_{c+}=\text{max}[I(0<\phi<2\pi)]$ and the critical current in the $-x$ direction is defined as $I_{c-}=\text{max}[-I(0<\phi<2\pi)]$. In the following, we present the numerical results of the JDE efficiency only for the chirality $\gamma=+1$. The JDE efficiency for the chirality $\gamma=-1$ is equal in magnitude but opposite in sign to that for $\gamma=+1$. This follows directly from Eq.(\ref{sr2}). In other words, the sign of the JDE efficiency can be tuned by the chirality of the HM in our junctions. In experiment, the control of chirality in helimagnetic materials using electric current pulses has been demonstrated\cite{Masuda2}.

In Fig.\ref{fig3}, we present the JDE efficiency as a function of the tilt angle $\alpha$ and the exchange coupling strength $\lambda$. From Figs.\ref{fig3}(a)-(c), we observe that the JDE efficiency $\eta$ oscillates with the tilt angle. For $\alpha=0$ or $\pi$, the efficiency is exactly zero as shown in Figs.\ref{fig3}(a)-(c), which follows from Eq.(\ref{sr1}). In this case, the HM is in the helical magnetic state and the nonreciprocity of the Josephson supercurrent vanishes. For $\alpha=\frac{\pi}{2}$, the efficiency is also zero as shown in Figs.\ref{fig3}(a)-(c) due to the relation in Eq.(\ref{srpi2}). In this case, the HM becomes a ferromagnet. These results are consistent with the CPRs presented in Fig.\ref{fig2}. Additionally, the JDE efficiency is antisymmetric about $\alpha=\pi$ as shown in Figs.\ref{fig3}(a)-(c). This property follows directly from Eq.(\ref{sr1}). Therefore, the necessary condition for JDE in our junctions is $\alpha\ne m\pi/2$, where $m$ is an integer.
Given that the diode efficiency $\eta$ must satisfy $\eta(\alpha) = -\eta(-\alpha)$ and vanish at the symmetry points $\alpha = m\pi/2$, this forces $\eta$ to complete one oscillation within each interval $[m\pi/2, (m+1)\pi/2]$, which is precisely the behavior observed in Figs. 3(a-c).

In Fig.\ref{fig3}(a), the JDE efficiency for different strengths of the exchange coupling $\lambda$ is presented. For $\lambda=0$, the HM becomes a normal metal with no coupling between electrons and the local magnetic moment and JDE does not occur as shown by the black line. Indeed, for small values of $\lambda$ such as $\lambda=0.4$, the nonreciprocity of the Josephson current is not pronounced as shown by the red line in Fig.\ref{fig3}(a). However, as $\lambda$ increases to $\lambda=0.8$, the nonreciprocity of the Josephson current becomes very strong and the efficiency of JDE can exceed $20\%$. For a larger value of $\lambda$ with $\lambda=1$, the maximum efficiency can approach $40\%$ as shown by the green line. A further increase in $\lambda$ suppresses the nonreciprocity of the Josephson supercurrent as illustrated by the purple line for $\lambda=1.2$.

In Fig.\ref{fig3}(b), the JDE efficiency for different values of the chemical potential in HM is presented. For $\mu=0$, JDE is absent in our junctions, regardless of the tilt angle $\alpha$. Even for a general $\alpha$ corresponding to the conical magnetic configuration, JDE does not occur. This can be understood from the energy bands in Fig.\ref{fig1}(c). For the conical magnetic configuration, although the energy bands are not symmetric about $k=0$, the relation $E(k)=-E(-k)$ still holds for the opposite spin projection\cite{Deaconu,Brekke}. In this case, the two electrons involved in pairing have the opposite momentum and the Cooper pairs with finite-momentum cannot be formed. Finite-momentum Cooper pairs are responsible for the realization of JDE and their absence in our junctions for $\mu=0$ explains why JDE does not occur. For a nonzero chemical potential, the relation $E(k)=-E(-k)$ is no longer satisfied at the Fermi surface and JDE emerges as shown in Fig.\ref{fig3}(b). Moreover, the maximum JDE efficiency initially increases and then decreases as the chemical potential increases. From the above discussion, we find that another necessary condition for JDE in our junctions is $\mu\ne0$.

In Fig.\ref{fig3}(c), the JDE efficiency for different numbers of supercells in HM is presented. As $N$ increases from $N=1$ to $N=10$, the maximum JDE efficiency does not decrease significantly. However, the oscillation in the JDE efficiency becomes more pronounced, because the HM chain becomes longer. With a longer HM chain, the phase acquired by carriers as they traverse the junction depends more sensitively on the parameters, which results in more pronounced oscillations.
For the practical applications, short junctions with $N\le 5$ are more stable against variations in $\alpha$. Indeed, helical-spin magnets with a short period spin-helix state have been realized experimentally\cite{Kitaori,Yokouchi2}, which makes our scheme for JDE practically feasible.

In Figs.\ref{fig3}(d)-(f), we present the variations of the JDE efficiency with the exchange coupling strength $\lambda$. As discussed above, JDE vanishes when $\lambda=0$, which is consistent with the result for $\lambda=0$ in Fig.\ref{fig3}(a). When $\lambda>1.5$, the JDE efficiency decreases and rapidly approaches zero, which is also consistent with the suppression of diode efficiency at large $\lambda$ in Fig.\ref{fig3}(a). However, for moderate $\lambda$ values, the JDE efficiency can approach $40\%$ for $\alpha=0.1\pi$ or $0.3\pi$ (Fig.\ref{fig3}(d)), $30\%$ for $\mu=1$ (Fig.\ref{fig3}(e)) and also $30\%$ for $N=1$ (Fig.\ref{fig3}(f)). Furthermore, high efficiency can be achieved in a broad range of $\lambda$, which demonstrates the robust feasibility of the diode effect in our junctions.

The realization of JDE and its fundamental characteristics are closely linked to both the broken and preserved symmetries of the quantum materials in the Josephson junctions. From the above discussion, it can be concluded that the vanishing of the diode efficiency and its antisymmetric behavior directly reflect the symmetric properties of HM and SCs. These features are independent of the detailed parameters of the junctions and can be well explained through the symmetry analysis. However, the detailed dependence of the diode efficiency on the parameters is too complex to admit a simple explanation. In fact, the oscillatory behavior as a function of parameters is a typical characteristic of diode efficiency in various Josephson structures\cite{Costa1,Mondal,CSun,YFSun2,Vakili} and is not unique to helimagnetic Josephson junctions.

We now briefly discuss the difference between our scheme and the existing studies. First, the realization of JDE here does not require spin-orbit coupling, which is distinct from the JDE based on the Rashba spin-orbit coupling\cite{Maiani,Costa1,WTLu,Mondal,Ilic,Cheng2,Debnath,YMao,Cayao,WTLiu}. Second, our junctions consist of a simple 1D HM chain\cite{Deaconu} and two conventional $s$-wave SCs, in contrast to the two-dimensional junctions and the three-dimensional junctions with helical magnets recently studied in Refs.[\onlinecite{Nikolic}] and [\onlinecite{Kamra}]. In the former, superconductivity is proximity-induced in the magnet. In the latter, strongly spin-polarized ferromagnetic trilayers with a conical magnetic configuration are considered. Compared to these two types of systems, our junctions possess a simpler structure and a clearer physical picture, which aids in understanding the physical mechanism of JDE in the helimagnetic Josephson junctions.

In our calculations, we have chosen $t=1$, $M=1$ and $\lambda\in[0,2]$, which are consistent with the parameter selection in Ref.[\onlinecite{Deaconu}]. In Ref.[\onlinecite{Deaconu}], $t=1$, $M=1$ and $\lambda=0.6$ or $0.8$ were chosen and these parameters effectively capture the essential physics underlying the band asymmetry and the nonreciprocal transport observed in $\alpha$-EuP$_3$\cite{Mayo}. Furthermore, the parameter choices in Ref.[\onlinecite{Deaconu}] are directly relevant to the real material $\alpha$-EuP$_3$ with the hopping amplitude $t=8meV$, the coupling strength $\lambda$ of several $meV$ and the local magnetization $M=7\mu_{B}$(see Supporting Information of Ref.[\onlinecite{Deaconu}]). Therefore, it is expected that the JDE proposed in this paper can be realized in the SC/$\alpha$-EuP$_3$/SC junctions. Additionally, the diode efficiency of about $5\%$ can be observed in experiment and a high rectification ratio can be achieved\cite{HWu}. It is believed that the diode efficiency in our junctions should be experimentally observable over a wide range of parameters near those yielding the maximum efficiency.

Finally, we discuss JDE in our junctions when the $s$-wave SCs are replaced by the spin-triplet $p_x$-wave SCs. Numerical calculations show that the relations in Eqs. (\ref{sr1}), (\ref{srpi2}) and (\ref{sr2}) still hold for the $p$-wave junctions. For the helical configuration with $\alpha=0$ and the ferromagnetic configuration with $\alpha=\frac{\pi}{2}$, JDE does not occur in the $p$-wave junctions. The diode efficiency remains antisymmetric about $\alpha=\pi$ and its sign can be reversed by changing the chirality of the HM. These basic characteristics are the same as those for the $s$-wave junctions. However, there is an important difference between the two types of junctions. For the $p$-wave junctions, the diode efficiency does not vanish when $\mu=0$. As discussed above, for electrons with the opposite spin projection at $\mu=0$, the energy bands satisfy $E(k)=-E(-k)$\cite{Deaconu,Brekke}, preventing the formation of finite-momentum pairing in the $s$-wave junctions. However, for electrons with the same spin projection at $\mu=0$, the energy bands do not satisfy $E(k)=-E(-k)$. The tunneling of the equal-spin Cooper pairs becomes possible for the conical magnetic configuration in the HM when the SCs have the $p$-wave pairing. Consequently, finite-momentum equal-spin pairing and thus JDE mediated by the equal-spin Cooper pairs can occur at $\mu=0$ for the $p$-wave junctions.

\section{\label{sec4}Conclusions}
The nonreciprocity of the Josephson supercurrent is studied in the 1D SC/HM/SC junctions in the absence of spin-orbit coupling. JDE is absent for the helical magnetic and ferromagnetic configurations of the HM but emerges for the conical magnetic configuration with a finite chemical potential in the $s$-wave junctions. The magnitude and sign of the diode efficiency can be controlled by tuning parameters such as the chirality, tilt angle, exchange coupling, chemical potential and size of the HM. The mechanism and the characteristics of the nonreciprocity in our junctions have been analyzed through Hamiltonian transformations and energy band calculations. The supercurrent nonreciprocity in the $p$-wave junctions is also investigated. The nonzero diode efficiency can be achieved even for zero chemical potential due to the tunneling of the equal-spin Cooper pairs. The high diode efficiency achieved in our system may facilitate the design of low-power-consumption electronic devices.

\section*{\label{sec5}ACKNOWLEDGMENTS}

This work was financially supported
by the National Natural Science Foundation of China
under Grants Nos. 12474046, 12374034 and 12447146,
the National Key R and D Program of China (Grant No. 2024YFA1409002),
the Quantum Science and Technology-National Science and
Technology Major Project (2021ZD0302403),
the projects ZR2023MA005 and ZR2022QA110 supported by Shandong Provincial Natural Science Foundation,
the China Postdoctoral Science Foundation (No. 2025T180938),
and the Postdoctoral Fellowship Program of CPSF under Grant No. GZB20240031. We acknowledge
the High-performance Computing Platform of Peking University
for providing computational resources.

\appendix

\section{\label{secA} \uppercase{The calculation of the energy band in HM}}
\setcounter{equation}{0}
\setcounter{figure}{0}
\renewcommand{\theequation}{A\arabic{equation}}
\renewcommand{\thefigure}{A\arabic{figure}}
The four atoms in each supercell of the HM are labeled by $A$, $B$, $C$ and $D$ from left to right.
After a Fourier transform, the tight-binding Hamiltonian of the HM in Eq.(\ref{hHM}) can be written as
\begin{eqnarray}
h=\sum_{k}\psi_{k}^{+}h_{k}\psi_k
\end{eqnarray}
with $\psi_{k}=(c_{Ak\uparrow},c_{Ak\downarrow},c_{Bk\uparrow},c_{Bk\downarrow},c_{Ck\uparrow},c_{Ck\downarrow},c_{Dk\uparrow},c_{Dk\downarrow})^{T}$ and
\begin{eqnarray}
\begin{split}
&h_{k}=\\
&\left(\begin{array}{cccccccc}
A_{\uparrow\uparrow}&A_{\uparrow\downarrow}&x_k&0&0&0&x_k^{*}&0\\
A_{\downarrow\uparrow}&A_{\downarrow\downarrow}&0&x_k&0&0&0&x_k^{*}\\
x_k^{*}&0&B_{\uparrow\uparrow}&B_{\uparrow\downarrow}&x_k,&0&0&0\\
0&x_k^{*}&B_{\downarrow\uparrow}&B_{\downarrow\downarrow}&0&x_k&0&0\\
0&0&x_k^{*}&0&C_{\uparrow\uparrow}&C_{\uparrow\downarrow}&x_k&0\\
0&0&0&x_k^{*}&C_{\downarrow\uparrow}&C_{\downarrow\downarrow}&0&x_k\\
x_k&0&0&0&x_k^{*}&0&D_{\uparrow\uparrow}&D_{\uparrow\downarrow}\\
0&x_k&0&0&0&x_k^{*}&D_{\downarrow\uparrow}&D_{\downarrow\downarrow}
\end{array}\right),\label{hHMk}
\end{split}
\end{eqnarray}
with $x_{k}=-t_0 e^{jka}$ and
\begin{eqnarray}
\begin{split}
A_{\uparrow\uparrow}&= \lambda M\cos{\alpha}-\mu,&
A_{\uparrow\downarrow}&=\lambda M \sin{\alpha},\\
A_{\downarrow\downarrow}&= -\lambda M\cos{\alpha}-\mu,&
A_{\downarrow\uparrow}&=\lambda M \sin{\alpha},\\
B_{\uparrow\uparrow}&=-\mu,&
B_{\uparrow\downarrow}&= \lambda M(\sin{\alpha}-j\gamma\cos{\alpha}),\\
B_{\downarrow\downarrow}&=-\mu,&
B_{\downarrow\uparrow}&= \lambda M(\sin{\alpha}+j\gamma\cos{\alpha}),\\
C_{\uparrow\uparrow}&= -\lambda M\cos{\alpha}-\mu,&
C_{\uparrow\downarrow}&=\lambda M \sin{\alpha},\\
C_{\downarrow\downarrow}&= \lambda M\cos{\alpha}-\mu,&
C_{\downarrow\uparrow}&=\lambda M \sin{\alpha},\\
D_{\uparrow\uparrow}&=-\mu,&
D_{\uparrow\downarrow}&= \lambda M(\sin{\alpha}+j\gamma\cos{\alpha}),\\
D_{\downarrow\downarrow}&=-\mu,&
D_{\downarrow\uparrow}&= \lambda M(\sin{\alpha}-j\gamma\cos{\alpha}).\\
\end{split}\nonumber
\end{eqnarray}
Diagonalization of $h_k$ yields the single-particle energies in the HM. For small $k$ near the $\Gamma$ point, $x_k$ can be expanded as $x_k\approx-t_0(1+jka)$. Taking $\lambda=1$, $M=1$, $t_0=1$, $\gamma=+1$ and $\mu=0$, the single-particle energies are given by
\begin{eqnarray}
E_{1}=1-k-\sqrt{k^2+2k+2-2(1+k)\sin{\alpha}},
\end{eqnarray}
\begin{eqnarray}
E_{2}=1-k+\sqrt{k^2+2k+2-2(1+k)\sin{\alpha}},
\end{eqnarray}
\begin{eqnarray}
E_{3}=1+k-\sqrt{k^2-2k+2-2(k-1)\sin{\alpha}}
\end{eqnarray}
\begin{eqnarray}
E_{4}=1+k+\sqrt{k^2-2k+2-2(k-1)\sin{\alpha}},
\end{eqnarray}
\begin{eqnarray}
E_{5}=-1-k-\sqrt{k^2-2k+2+2(k-1)\sin{\alpha}},
\end{eqnarray}
\begin{eqnarray}
E_{6}=-1-k+\sqrt{k^2-2k+2+2(k-1)\sin{\alpha}},
\end{eqnarray}
\begin{eqnarray}
E_{7}=-1+k-\sqrt{k^2+2k+2+2(k+1)\sin{\alpha}},
\end{eqnarray}
\begin{eqnarray}
E_{8}=-1+k+\sqrt{k^2+2k+2+2(k+1)\sin{\alpha}},
\end{eqnarray}
which correspond to the energy bands in Figs.\ref{fig1}(b)-(d). For $\gamma=-1$, the energy bands satisfy the relation $E(\gamma=-1,k)=E(\gamma=+1,-k)$ and the explicit expressions are not shown here.\\

\section{\label{secB}  \uppercase{The representations of operators }}
Here, we provide the representations of operators in Eqs.(\ref{eqX}),(\ref{Myztr}) and (\ref{eqZ}) through their actions on the annihilation operators $c_{i\uparrow}$ and $c_{i\downarrow}$ of electrons. For the time-reversal operator $T$, we have
\begin{eqnarray}
\begin{split}
T c_{i\uparrow}T^{-1}&=c_{i\downarrow},\\
T c_{i\downarrow}T^{-1}&=-c_{i\uparrow}.
\end{split}
\end{eqnarray}
For the gauge transformation $U_{1}(\frac{\pi}{2})$, we have
\begin{eqnarray}
\begin{split}
U_{1}(\frac{\pi}{2}) c_{i\uparrow}U_{1}(\frac{\pi}{2})^{-1}&=c_{i\uparrow}e^{i\frac{\pi}{2}},\\
U_{1}(\frac{\pi}{2}) c_{i\downarrow}U_{1}(\frac{\pi}{2})^{-1}&=c_{i\downarrow}e^{i\frac{\pi}{2}}.
\end{split}
\end{eqnarray}
For the mirror reflection $M_{xz}$ about the $xz$ plane, we have
\begin{eqnarray}
\begin{split}
M_{xz} c_{i\uparrow}M_{xz}^{-1}&=c_{i\downarrow},\\
M_{xz} c_{i\downarrow}M_{xz}^{-1}&=c_{i\uparrow},
\end{split}
\end{eqnarray}
and for the mirror reflection $M_{yz}$ about the $yz$ plane, we have
\begin{eqnarray}
\begin{split}
M_{yz} c_{i\uparrow}M_{yz}^{-1}&=c_{-i\downarrow},\\
M_{yz} c_{i\downarrow}M_{yz}^{-1}&=c_{-i\uparrow}.
\end{split}
\end{eqnarray}
Using these operators, we construct the joint operations $X$, $Y$ and $Z$ in Eqs.(\ref{eqX}),(\ref{Myztr}) and (\ref{eqZ}) and derive the relations satisfied by the supercurrent and the diode efficiency.

\end{document}